# Low energy structures in nuclear reactions with 4n in the final state

[1]Rimantas Lazauskas

[1]*Université de Strasbourg, CNRS, IPHC UMR 7178, F-67000 Strasbourg, France*

[2,3]Emiko Hiyama

[2]*Department of Physics, Tohoku Univ., Sendai 980-8578, Japan and*
[3]*RIKEN Nishina Center, Wako 351-0198, Japan*

[4]Jaume Carbonell

[4]*Université Paris-Saclay, CNRS/IN2P3, IJCLab, 91405 Orsay, France*

(Dated: February 15, 2023)

We investigate a reaction model that describes a fast removal of the $\alpha$-particle from the $^8$He nucleus with eventual emission of four neutrons. The obtained four neutron energy distribution allows one to explain the sharp low energy peak observed by studying the missing mass spectra of four neutrons in [Nature Vol. 606, p. 678], as a consequence of dineutron-dineutron correlations. The phenomenon of the emergence of a sharp low-energy peak in the four-neutron energy distribution should be more general and is expected in the decay of other systems containing four-neutron halo.

The possible existence of $3n$ and $4n$ bound and/or resonant states has been considered since the 60's. The interest in this topic was, however, boosted at the beginning of this century by the experimental findings at GANIL [1, 2] and at RIKEN [3] claiming a positive signal of a near-threshold bound or resonant tetraneutron.

While the existence of bound three- or four-neutron states is totally excluded by theorists [4–6], as well as by most of experimental studies [7], there remains a debate, both from the theoretical and experimental points of view, about the existence of multineutron resonances [8].

In our opinion, the possibility of observing $3n$ or $4n$ resonances, a fortiori bound states, is also excluded. This is a direct and model independent consequence of Effective Field Theory (EFT) in the unitary limit [9–11], which predicts strong repulsion between two identical difermions in the total angular momentum $J = 0^+$ state, the most favorable configuration to form four fermion states [12–14]. The same negative conclusion is shared by a series of theoretical studies, implementing properly the asymptotic behavior of unbound systems [4, 12–25]. Inline with the EFT arguments, the conclusions of the former studies were found to be independent from the details of the $nn$ interaction, and not affected by the presence of a realistic three-nucleon force. These findings are in sharp contrast with the paradoxical results of Refs. [6, 26–28] in which the presence of any $4n$ bound state is largely excluded but the possibility of a $4n$ near-threshold resonance is suggested. According to [21, 25], this discrepancy is due to unsound extrapolation procedure from the bound state region to the continuum [6, 27, 28], but also a consequence of disregarding the long-range dineutron correlations [26, 28], a crucial component of the multineutron dynamics [22].

Nevertheless, several studies [12–18, 23] indicate that sharp low energy structures might be formed in the four-neutron production cross section as an interplay of multineutron dynamics and a complex reaction mechanism, without any link to a $4n$ resonant state. A trivial illustrative example can be found in Fig. 21 of Ref. [8] where such structures are produced by a repulsive well potential.

A recent publication [29] reports evidence of a low energy structure in the quasi-elastic reaction $^8$He($p$, $p^4$He)$4n$ performed at RIKEN. This remarkable study provides the first convincing signal of a near-threshold structure in a nuclear reaction with 4 neutrons in the final state. The signal, composed of two well-separated peaks at E≈2 MeV and at E≈30 MeV respectively, is observed in the missing mass spectrum of the $4n$ system. Assuming a Breit-Wigner form, it is suggested that the first peak could correspond to a $4n$ resonance with parameters $E_R$=2.37±0.38±0.44 MeV and $\Gamma$=1.75±0.22±0.30 MeV. The analysis, based on the COSMA model [30] which assumes a sudden removal of the alpha-particle core from $^8$He and a flat distribution of the four neutron final state, provided an almost perfect description of the broad structure, but found no explanation for the sharp low energy peak. However, the authors of Ref. [15], employing the same model, observed a strong dependence of the multineutron response on their initial distribution and final state interactions, in shifting the peak to lower energies. The last study was however unable to consider in full extent the four-neutron correlations generated by $2n+2n$ configurations.

Here we aim to build a realistic model of the $^8$He($p$,$p^4$He)$4n$ reaction to explain the low energy structure reported in [29], bridging the gap between the conflicting views in theory and experiment. We show that these experimental results find a natural explanation in terms of the dineutron correlations in the final state, if the four neutrons are weakly bound in the initial projectile, forming a broad wave function.

The kinematical conditions of the $^8$He($p$, $p^4$He)$4n$ reaction in [29], are such that the bulk of the 156 MeV/nucleon kinetic energy carried by the $^8$He projectile is transferred from the $\alpha$-particle – constituting the core of the $^8$He – to the proton. In the center-of-mass frame of the $^8$He nucleus, this translates into a sudden removal of the $\alpha$-particle followed by an eventual dissipation of the four slow valence neutrons. Obviously, the initial distribution of these valence neutrons plays an important role in determining their break-up profile. This feat was observed already in [15], whose model turned out to be very successful in describing the high energy part of the $4n$ distribution in [29]. In the latter analysis, however, the four valence neutrons in $^8$He were considered to be filling up the lowest symmetry allowed harmonic oscillator (HO) shells. Such approach mimics well the exchange of the valence neutrons with those present in the $\alpha$-particle core, but strongly overestimates the kinetic energy and fails to describe the complexity of the $^8$He wave function. No Core Shell model results reveal that within $0\hbar\omega$ model space, the four valence neutrons are strongly repelled from the $^4$He core [32–34]. Very large HO model space is required to make the $^8$He nucleus bound relative to the $\alpha$-particle and even larger space to bind it relative to the $^6$He ground state, thus suggesting an important clustering and correlations of the valence neutrons in $^8$He.

We complement the analysis of the $^8$He$(p,p^4$He$)4n$ reaction, by addressing the aforementioned shortcomings of the COSMA model [29] in three essential ways: ($i$) implementing a realistic description of the $^8$He valence neutron distribution, ($ii$) implementing a rigorous dynamics for the four-neutron break-up, and ($iii$) considering the interaction between valence neutrons in full extent and retaining consistency between the multineutron Hamiltonians before and after the $\alpha$-particle removal.

The valence neutron distribution in $^8$He ($^6$He) is simulated by a Hamiltonian describing the four (two) neutrons in the mean field created by the $\alpha$-core, assumed to coincide with the center-of-mass of the valence neutrons $\vec{R}_G = \frac{1}{N}\sum_{j=1}^{N}\vec{r}_j$. The initial Hamiltonian, prior to $\alpha$-particle removal, is given by

$$H_i = H_0 + \lambda\sum_{i=1}^{N}|\psi_\alpha(r_i)\rangle\langle\psi_\alpha(r_i)| + \sum_{i<j=1}^{N}V_{nn}(r_{ij}) + \sum_{i=1}^{N}V(r_{iG}) + \sum_{i<j=1}^{N}W_{ij}(\rho, r_{ijG}), \quad (1)$$

where $H_0$ is the kinetic energy of the N=2(4) valence neutrons, $V_{nn}$ is the neutron-neutron interaction depending on the interparticle distance $\vec{r}_{ij} = |\vec{r}_i - \vec{r}_j|$,

$$V(r_{iG}) = V_0\, e^{-[(\vec{r}_i - \vec{R}_G)/\rho_0]^2}, \quad (2)$$

is the mean-field acting on each neutron with position $\vec{r}_i$.

$$W_{ijG}(\rho, r_{ijG}) = r_{ijG}\, W_0 e^{-\left(\frac{\rho}{\rho_0}\right)^2}; \quad \rho^2 = \frac{r_{ij}^2}{16} + \frac{r_{ijG}^2}{2} \quad (3)$$

is a three-body force between two neutrons and the $^4$He core, with $\vec{r}_{ijG} = (\vec{r}_i + \vec{r}_j)/2 - \vec{R}_G$. To account for the Pauli repulsion between the valence and the core neutrons we employ the projection method [31]. A nonlocal term $|\psi_\alpha(r_i)\rangle\langle\psi_\alpha(r_i)|$ is introduced in (1) where $|\psi_\alpha(r_i)\rangle$ is the ground state wave function of the HO, with the oscillator parameter chosen to the experimental rms radius of the $\alpha$-particle $\langle r_n^2\rangle^{\frac{1}{2}}(^4\text{He})$=1.45 fm. The projection parameter $\lambda$ should be large, ideally $\lambda \to \infty$.

The model parameters $(\rho_0, V_0, W_0)$ are determined as follows. For some selected $\rho_0$ values, the strength parameters $V_0$ of the mean field (2) is adjusted to the two-neutron separation energy of $^6$He: $S^{2n}_{6He} \equiv$ B($^6$He)-B($^4$He)=0.97 MeV. Notice that for $^6$He, one has by construction $r_{ijG} = 0$ and consequently the three-body force (3) does not contribute. Its strength parameter $W_0$ is then adjusted to the $4n$ separation energy of $^8$He: $S^{4n}_{8He} \equiv$ B($^8$He)-B($^4$He)=3.11 MeV.

The initial state $|\Psi_i\rangle$, representing the $^8$He nucleus, is provided by a solution of the Schrödinger equation

$$H_i \,|\, \Psi_i\rangle = E_i\,|\,\Psi_i\rangle. \quad (4)$$

It has been obtained by solving the corresponding Faddeev-Yakubovsky (FY) equations [36].

We further assume that the $^8$He projectile is broken by a sudden removal of the $^4$He-core. Thus, in the initial state, the four-neutron distribution coincides with that of the $^8$He valence neutrons. The four-neutron wave function is driven by the Hamiltonian:

$$H_f = H_0 + \sum_{i<j=1}^{4}V_{nn}(r_{ij}). \quad (5)$$

The final state, $|\Psi_f\rangle$, corresponds to the $4n$ in the continuum with total energy $E_{4n}$, and is a solution of

$$H_f|\Psi_f\rangle = E_{4n}|\Psi_f\rangle. \quad (6)$$

Next step is to obtain the response (or strength) function $S(E)$ corresponding to the process $\langle^4\text{He}\Psi_{4n}(E)|\hat{O}|^8\text{He}\rangle$, where $\hat{O}$ is a transition operator representing the effect of $\alpha$-core removal on the initial configuration of the four valence neutrons. If the $\alpha$-core is removed without affecting the peripheral neutrons in the halo $\hat{O} = 1$. As explained in [18] the response function is given:

$$S_{4n}(E) = -\frac{1}{\pi}\,\text{Im}\left\langle \Psi_i \left| \hat{O}^\dagger \right| \Psi_f^+(E)\right\rangle, \quad (7)$$

where the wave function $\Psi_f^+(E)$ is a solution of the inhomogeneous equation

$$(E - H_f + i\epsilon)\Psi_f^+(E) = \hat{O}\Psi_i, \quad (8)$$

at a chosen energy E.

The right-hand side of (8) is square-integrable, damped by the bound-state wave function $\Psi_i$. $\Psi_f^+$ contains asymptotically only outgoing waves with a rather complicated structure, involving multidimensional four-neutron break-up amplitudes. Nevertheless the last inhomogeneous equation may be comfortably solved using the complex scaling technique, as explained in [18, 42]. Notably, the numerical calculations are realized using the same techniques as described in our former work [36].

In conjunction with the phenomenological interaction (2) and (3) we have considered three different $nn$ potentials: AV18 [37], providing the low energy parameters $a_{nn}$=-18.8 fm and $r_0$=2.83 fm, $\chi$N3LO potential [38] based on chiral-EFT, providing $a_{nn}$=-18.9 fm and $r_0$ =2.84 fm and MT13 S-wave interaction [39] adjusted to $a_{nn}$=-18.6 fm and $r_0$=2.93 fm. As pointed out in [12, 13, 16–18] none of the aforementioned Hamiltonians support a 4n near-threshold resonant states that could generate a low energy peak. Furthermore, MT13 providing $a_{nn}$ value much larger than the range of the nuclear interaction, fully complies with the EFT predictions in the unitary limit [9–11], which indicates the absence of an attractive interaction between two resonant fermionic pairs.

To test our model, we computed for selected $(V_0, W_0)$ values reproducing the experimental neutron separation energies ($S^{2n}_{^4He}$=0.97 MeV and $S^{4n}_{^4He}$=3.11 MeV), the neutron rms radii $\langle r_n^2 \rangle_{^AHe}$ of the $^6$He and $^8$He ground states. The AV18 results are listed in Table I. These radii are estimated from the calculated rms radii of the valence neutrons $\langle r_n^2 \rangle_{val}$, as

$$(A-2)\langle r_n^2 \rangle_{^AHe} \approx (A-4)\langle r_n^2 \rangle_{val} + 2\langle r_n^2 \rangle_{^4He}, \quad (9)$$

with A=6,8. Results of Table I show that the value $\rho_0$=2.5 fm represents a good compromise for describing both $^{6,8}$He isotopes. The four valence neutrons slightly deform the $^4$He core by attracting core-protons and compressing core-neutrons, what should lead to slightly smaller neutron rms radii than those estimated by equation (9).

The four neutron strength function (7) has been computed for $\rho_0$=2.5 fm, and it is displayed in the upper panel of Figure 1 for three different choices of $V_{nn}$: AV18 as cross symbols, MT13 as empty blue circles, $\chi$N3LO as empty red up-triangles. Independently of the $nn$ interaction we obtain a pronounced low energy peak for the four-neutron missing mass distribution centered at around 2.5 MeV. The model dependence is less than 2% at the peak and is only slightly visible at higher energy. The addition of a three-neutron term (3NF), to the $\chi$N3LO potential [40], has no visible effect on the 4n distribution. This model independence implies a negligible effect of $L > 0$ partial waves in $V_{nn}$, as expected from EFT.

To compare our calculations with those of Ref. [29], we have broadened our neutron strength function with the experimental resolution of 2 MeV and convoluted it with the experimental acceptance. These results were normalized to be consistent with the observed 54 events in the neutron missing-mass window $E_{4n} < 10$ MeV (see Fig 3 from [29]). Our results with AV18 and several values of $\rho_0$ are shown in the lower panel of Fig. 1.

We were able to obtain a reasonable description of the low energy peak centered at around $E_{4n} \approx$2.5 MeV, and that independent of the range parameter of the phenomenological model for the valence neutron distribution in $^8$He. The centroid and widths of the peaks correlate strongly with the rms radii of the neutron initial distribution, being pushed to lower energy if a more peripheral neutron distribution is generated in the initial state. This is in concordance with an observation made in [15]. However our distributions are much sharper and are centered at substantially lower energies, indicating the key importance of the 2n+2n decay channels, which were not fully considered in [15].

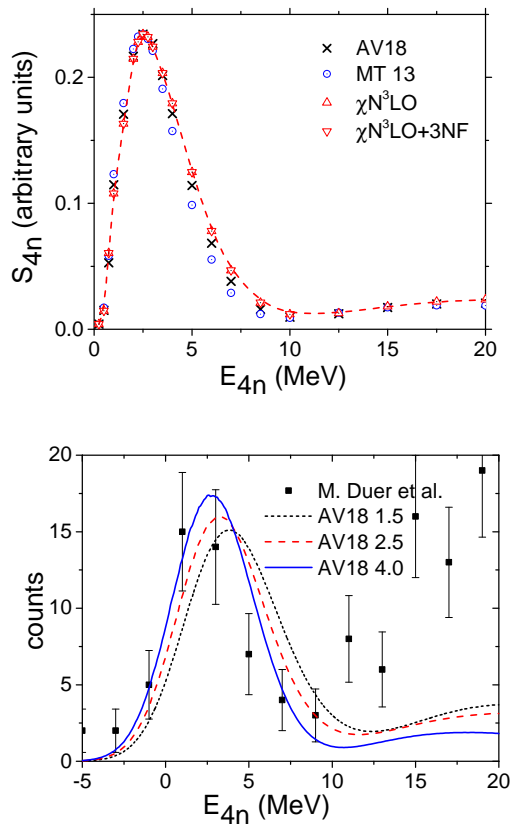

FIG. 1: Upper panel: dependence of the strength function $S_{4n}$ on $E_{4n}$ for the range parameter $\rho_0$=2.5 fm. We have used several $nn$ potentials: AV18 (cross symbols), MT13 (blue empty circles), $\chi$N3LO (red empty up-triangles). They all show a pronounced peak at $E_{4n} \approx$2.5 MeV and the model dependence is very weak ($\leq 2\%$). We added a three-neutron force to $\chi$N3LO (red empty inverted-triangles) with no significant effect. Lower panel: the strength function, broadened with the experimental resolution of 2 MeV and convoluted with the experimental acceptance, is compared to the measurement of [29]. Calculations, corresponding different $\rho_0$, are performed for AV18 $nn$ interaction and normalized to the number of observed counts below $E_{4n}$=10 MeV.

Noteworthy that the best agreement with the experimental data of [29] is obtained with $\rho_0 = 4$ fm, which however overestimates the neutron point rms radii of $^6$He and $^8$He (see Table I). A more detailed analysis of these neutron distributions reveals that the interaction range $\rho_0$ mostly acts in separating 2n+2n clusters in the wave function of $^8$He. Our $^8$He model considerably simpli-fies the Pauli principles action between the $^4$He-core and the four valence neutrons, which strongly enhances the 2n+2n cluster separation as we have observed by comparing our results with those neglecting Pauli forbidden states. Moreover, one may expect that the full reaction mechanism amplifies the contribution of the peripheral neutrons to the low energy part of the response function. Actually, the neutrons staying close to the core are more energetic and correlate stronger with the core. They may gain some momenta with the core removal, propelling their contribution into the high energy peak of the missing mass spectra.

| $\rho_0$ (fm) | $V_0$ (MeV) | $W_0$ (MeV fm$^{-1}$) | $<r_n^2>^{\frac{1}{2}}$ ($^6He$) (fm) | $<r_n^2>^{\frac{1}{2}}$ ($^8He$) (fm) |
|---|---|---|---|---|
| 1.5 | -118.60 | -2.553 | 2.55 | 2.92 |
| 2.5 | -61.757 | -0.2125 | 2.66 | 3.05 |
| 4.0 | -22.114 | -0.0507 | 3.12 | 3.72 |
| | | | 2.90(8) [41], 2.72(7) [35] | 2.92(4) [41], 2.67(7) [35] |

TABLE I: Strength parameters of the interactions (2) and (3) as a function of the interaction range $\rho_0$, adjusted to reproduce the experimental $^6$He and $^8$He neutron separation energies. In the two last columns, the corresponding calculated neutron rms radii are compared with an estimation from the experimental (p,p') scattering data [35] and *ab-initio* calculation [41].

Up to this point we have considered that the removal of the $\alpha$−particle from $^8$He is instantaneous and that this process leaves the valence neutrons unaffected in the total angular momentum $J^\pi = 0^+$ state. This is certainly a very good approximation, which proved to be successful even in describing the high energy parts of the two- and four-neutron response in [29], respectively measured for the $^6$He and $^8$He decays. For the sake of completeness, we have simulated the effect of the core-recoil corrections, provided by simple transition operators $\hat{O}$ in (7). We have considered a set of symmetry allowed spin-orbit operators, having the form $\hat{O} = \sum_i \{\vec{r}_i \otimes \vec{\sigma}_i\}_g$, and delivering a transition to four-neutron final states with $J^\pi = g^-$. It turns out that these transition operators generate remarkably similar distributions, which almost coincide when weighted by the statistical factor $1/(2g+1)$ (see Fig. 2). Relative to the unperturbed scenario $\hat{O} = 1$, the low energy peak is shifted to higher energy and becomes broader. This seems to be consequence of the former operator imposing a spin-flip and thus breaking the configurations where two resonant $^1S_0$ dineutron pairs are present. On the contrary, the choice $\hat{O} = \sum_i r_i^2 Y_2(\hat{r}_i)$, imposing a final four-neutron configuration with $J^\pi = 2^+$ and allowing break-up into two $^1S_0$ dineutrons with relative angular momentum L=2, results in an even sharper low energy peak in the four-neutron distribution than does $\hat{O}$=1. This feature emerges regardless the fact that, with the transfer of angular momentum L=2, the four neutrons necessarily gain rotational energy; this effect is nevertheless largely compensated by the emphasized contribution of more peripheral valence neutrons via the factor $r_i^2$.

In order to understand better the emergence of a low-energy peak in the four-neutron missing mass distribu-

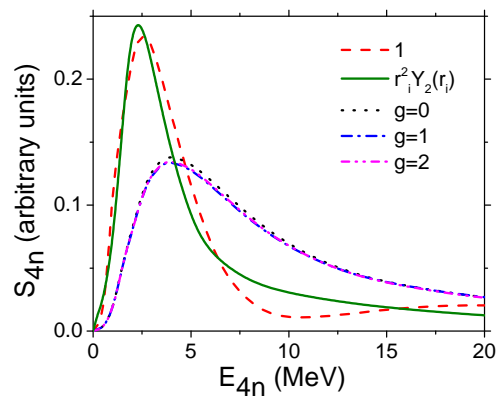

FIG. 2: Low energy four-neutron response functions calculated with the AV18 $nn$ interaction and $\rho_0$=2.5 fm. Different transition operators $\hat{O}$ were considered in order to visualize the effect of the core-recoil corrections. The olive-dashed curve corresponds to $\hat{O} = \sum_i^4 r_i^2 Y_2(\hat{r}_i)$; red-dotted together with the dashed-dotted curves to $\hat{O} = \sum_i \{\vec{r}_i \otimes \vec{\sigma}_i\}_g$, solid-black curve represents the reference result with $\hat{O}$=1.

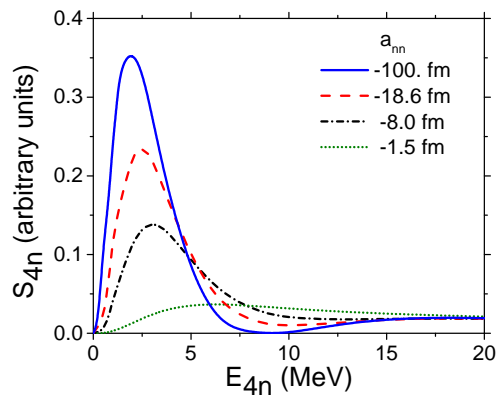

FIG. 3: Low energy 4n response functions for the scaled $nn$ MT13 potential adjusted to reproduce $a_{nn}$.

tion, we studied the impact of the dineutron-dineutron correlations. To accomplish this we have readjusted the MT13 potential to different $nn$ scattering length values: $a_{nn}$=-100, -8 and -1.5 fm. This was achieved by scaling the potential by a factor $\gamma$=1.0808, $\gamma$=0.89135 and

$\gamma=0.49005$ respectively. As previously, we have refitted the phenomenological mean-field interactions (2) (3) for $\rho_0=2.5$ fm in order to reproduce the proper $^6$He and $^8$He separation energies. These variations of $V_{nn}$ had very little effect on the calculated rms radii of the valence neutrons, changing them by only 1.5%. On the contrary, the neutron energy distributions indicate a strong inverse correlation with the $a_{nn}$ size. As the $nn$ interaction approaches the unitary limit ($a_{nn} = \pm\infty$), the neutron energy distribution becomes more and more pronounced. However, if $a_{nn}$ becomes non-resonant (e.g. $a_{nn}$=-1.5 fm) the strength function completely flattens. Decreasing $a_{nn}$ from -18.6 fm to -8 fm displaces the $4n$ strength functions peak by roughly 0.5 MeV.

**In summary**, in a recent experiment performed in RIKEN [29] a remarkably sharp low energy structure was observed in the missing mass distribution of four neutrons emitted in the quasi-elastic knock-out reaction $^8$He(p,p$^4$He)4n. The authors of this experiment have not found an explanation for this phenomenon, though managing to describe successfully the $4n$ distribution at higher energies as well as the presence of a low energy signal in a similar decay of $^6$He. As a possible explanation, the existence of a low-energy four-neutron resonant state has been suggested, thus challenging the theoretical understanding of the four-neutron system.

Motivated by these astonishing experimental results, we have constructed a realistic reaction model to describe the sudden $\alpha$-particle removal from $^8$He. The model is based on a transition between the $^4$He+$4n$ initial state and the four interacting neutrons in the final one. A rigorous calculation allows us to determine the low energy distributions of the four remaining neutrons in the final state, which is in close agreement with the experimental data. In view of these results, we propose a natural explanation for the low energy structure observed in [29]: it emerges as a consequence of the final state interaction among the $4n$ and the – *important* – presence of four neutrons in the periphery of the $^8$He projectile.

Our calculations were constrained only by the requirement of four valence neutrons to be weakly bound by a nuclear core. Thus, our study addresses a class of reactions involving fast removal of the core from $4n$ halo nucleus, and reveals a non trivial phenomenon consisting in the emergence of a sharp low energy peak in the missing mass spectrum of a 4-neutron decay. Such phenomena might also be seen in some systems of cold atoms.

**Acknowledgments**. We are grateful to B.F. Gibson for his suggestions in preparing this manuscript. This research, started during the program Living Near Unitarity at the Kavli Institute for Theoretical Physics (KITP), University of Santa Barbara (California) that is supported in part by the National Science Foundation under Grant No. NSF PHY-1748958. We thank the organizers and the staff members of this Institute for their invitation and financial support. We have benefitted from the French IN2P3 for a theory project "Neutron-rich light unstable nuclei" and by the Japanese Grant-in-Aid for Scientific Research on Innovative Areas (No.18H05407). We also profited from enriching discussions during the ECT* NPES 2022 workshop in Trento and thank ECT* for partial support of RL and EH stay. We were granted access to the HPC resources of TGCC/IDRIS under the allocation A0110506006 made by GENCI (Grand Equipement National de Calcul Intensif).